\documentclass[pre, preprint, a4paper, amsmath, nofootinbib, floatfix] {revtex4}
\usepackage{graphicx}
\begin{document}
\def\s{\sigma}
\title{{\Large \bf Nonequilibrium Phase Transitions}\vspace{0.5truecm} \\
\it Lecture Notes, Les Houches, July 2002\vspace{1.5truecm}}
\vspace{1cm}
\noindent
\vspace{1cm}
\author{\Large \bf Zolt\'an R\'acz\vspace{0.5truecm}}

\affiliation{ \large Institute for Theoretical Physics, E\"otv\"os University,
P\'azm\'any s\'et\'any 1/a, 1117 Budapest, Hungary\vspace{2truecm}}
%

\begin{abstract}
\vspace{1truecm}
Nonequilibrium phase transitions are discussed with emphasis
on general features such as the role of detailed balance violation in
generating effective (long-range) interactions, the importance
of dynamical anisotropies, the connection between various
mechanisms generating power-law correlations, and the emergence of
universal distribution functions for macroscopic quantities. Quantum spin
chains are also discussed in order to demonstrate how to construct
steady-states carrying fluxes in quantum systems, and to explain
how the fluxes may generate power-law correlations.
\end{abstract}

\maketitle

\noindent {\bf  CONTENTS}
\bigbreak

\noindent \ref{intro}.   Introduction

\hspace{1truecm} Nonequilibrium steady states (NESS)

\hspace{1truecm} Problems with usual thermodynamic concepts
\bigbreak

\noindent \ref{ptffe}. Phase transitions far from equilibrium

\hspace{1truecm} Differences from equilibrium - constructing models with NESS

\hspace{1truecm} Generation of long-range interactions - nonlocal dynamics

\hspace{1truecm} Generation of long-range interactions - dynamical anisotropies

\hspace{1truecm} Driven lattice gases, surface growth

\hspace{1truecm} Flocking behavior
\bigbreak

\noindent \ref{powlaw}. Where do the power-laws come from?

\hspace{1truecm} Self-organized criticality (SOC)

\hspace{1truecm} Absorbing state transitions and their connection to SOC

\bigbreak

\noindent \ref{noneqdist}. Distribution functions in NESS

\hspace{1truecm} Power laws and universality of nonequilibrium distributions

\hspace{1truecm} Picture gallery of scaling functions

\hspace{1truecm} Upper critical dimension of the KPZ equation

\bigbreak

\noindent \ref{Quantumpt}. Quantum phase transitions

\hspace{1truecm} Spin chains with fluxes

\hspace{1truecm} Quantum effective interactions

\bigbreak

\noindent \ref{Outlook}. Outlook
\newpage

\section{Introduction}
\label{intro}

There are many reasons for studying nonequilibrium phase transitions.
Let us start be mentioning a few which carry some generality.

First and foremost, equilibrium in nature is more of an exception than the rule,
and structural changes (which constitute a significant portion of
interesting phenomena) usually take place in nonequilibrium conditions.
Thus there is much to be learned about
the complex ordering phenomena occurring far from equilibrium.

Second, while very little is understood about the general aspects of
nonequilibrium
systems, the equilibrium critical phenomena have been much studied and
have been shown to display universal features. This universality emerges
from large-scale fluctuations in such a robust way that one can expect
that similar mechanisms will work in nonequilibrium situations as well.
Thus, investigating the similarities and differences of equilibrium and
nonequilibrium orderings may help to discover the distinguishing
but still robust properties of nonequilibrium systems.

Third, power law correlations are present in many nonequilibrium phenomena
and there have been
many attempts to explain these correlations through general mechanisms.
Closer examination, however, usually reveals a close
connection to equilibrium or nonequilibrium critical phenomena.

My lectures were designed to revolve around problems related
to the above points.
The lectures are built on the theory of equilibrium phase transitions
\cite{{eq-pt},{HalpHoh}},
thus I assume knowledge about both
static and dynamic critical phenomena at least on the level of familiarity
with the basic concepts (symmetry breaking, order parameter,
diverging correlation length, order parameter, scaling, universality classes,
critical slowing down, dynamical symmetries).
I discuss simple examples throughout
so that enterprising students could try out their luck implementing
their ideas in simple calculations.
Due to space restrictions, however, not all of the details discussed
in the lectures and afterwards are covered in the written notes.
In particular, the picturesque parts of the explanations are often left out
since they take up disproportionally large part of the allowed space.
Nevertheless, these pictures may be important in both understanding
and memorizing, and I strongly
encourage the reader to go through the slides of the lectures, as well. They
can be found through the homepage of the school,
http://dpm.univ-lyon1.fr/houches\_ete/lectures/
or at http://poe.elte.hu/$\sim$racz/.

\subsection{Nonequilibrium steady states}
\label{noneqss}
\begin{figure}[hbtp]
	\centering
        \vspace{-0.8cm}
	\includegraphics[width=10.5truecm]{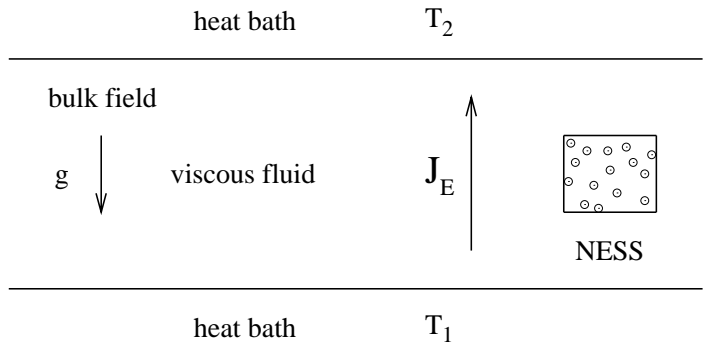}
        \vspace{-0.8cm}
	\caption[]{Setup for Rayleigh-B\'enard experiments.}
	\label{fig:Benard}
\end{figure}

A general feature that distinguishes a nonequilibrium steady state (NESS) from
an equilibrium one
is the presence of fluxes of physical quantities such as
energy, mass, charge, etc. Thus the study of NESS is, in a sense, a study
of the effects of fluxes imposed on the system
either by boundary conditions, or by bulk driving fields, or by some
combination of them. A nonequilibrium steady stateswell known example is shown on Fig.\ref{fig:Benard}.

This is the Rayleigh-B\'enard experiment \cite{Cross} in which
a horizontal layer of viscous fluid (the system) is heated from below i.e. it
has two heat baths of temperatures $T_1$ and $T_2$ attached (boundary conditions
generating an energy flux). The presence of gravity (the bulk drive)
is also important (it generates mass and momentum fluxes at large
$\delta T=T_1-T_2>0$).

For $T_1=T_2$ this system relaxes into a quiescent equilibrium state
while a small $\delta T$ will also result in a quiescent state but it
is already a NESS since energy flux is flowing through the system. Increasing
$\delta T$, this steady state displays a nonequilibrium phase transition
(Rayleigh-B\'enard instability), first to a stationary
convective pattern, and then to a series of more complicated structures
which have fascinated researchers for the past century \cite{Cross}.

Starting from the Navier-Stokes equations, one can arrive
at a mean-field level of understanding of the above phenomena. It is, however,
not the level of sophistication
one got used to in connection with equilibrium phase transitions.
There, we have simple exactly solvable models such as e.g. the Ising model
which give much insight into the mechanism of ordering and, furthermore,
this insight can be used to develop theories which reveal the
universal features of equilibrium orderings \cite{{eq-pt},{HalpHoh}}.

The trouble with the Rayleigh-B\'enard system is that we do not have a
theory even for the NESS. The reason for this is that the fluxes
result in steady state distributions , $P^*_n$, which break the
detailed balance condition $w_{n\to n'}P^*_n= w_{n'\to n}P^*_{n'}$, where
$n$ and $n'$ are two "microstates" and $w_{n \to n'}$ is the rate of
the $n \to n'$ transition. As a consequence of the breaking of detailed balance,
a NESS is characterized not only by the probability distribution, $P^*_n$,
but also by the probability currents in the phase space. Unfortunately, we
have not learned yet how to handle the presence of such loops of probability
currents.

The main lesson we should learn from the Rayleigh-B\'enard example
is that, in order to have Ising type models for describing
phase transitions in NESS, one should use models which relax to
steady states with fluxes present. Such models have been developed
during the last 20 years, and most of my lectures are about
these stochastic models defined through "microscopic"
elementary processes. The first level of description is in terms
of master equations which are conceptually simple and allow
one to make use of general results (uniqueness of stationary state, etc.)
which in turn are
helpful in defining dynamics that leads to NESS (Sec.\ref{ptffe}). The
next level is to describe the same problem in terms of Langevin
equations and develop field-theoretic techniques for the solution.
Our discussions will include both levels of description
and I hope that at the end an understanding will emerge
about a few results which grew in importance in the last decade
(generation of long-range interactions and effects of
dynamical anisotropies (Sec.\ref{ptffe}), connection between
mechanisms generating power-law correlations (Sec.\ref{powlaw}), and
universality of distribution functions for macroscopic quantities
(Sec.\ref{noneqdist})).

Before starting, however, I would like to insert here a little essay
about effective temperatures. This concept is being widely discussed in connection
with slowly relaxing systems, the topic of this school. So
it may be of interest to present here a view from the perspective of NESS.

\subsection{Problems with usual thermodynamic concepts}
\label{effT}
Systems close to equilibrium may retain many properties of an equilibrium
state with the slight complication that the intensive thermodynamic
variables (temperature, chemical potential, etc.) become inhomogeneous
on long lengthscales and they may slowly vary in time.
This type of situations are successfully dealt with
using the so called local equilibrium approximation \cite{loceq}, with the name
giving away the essence of the approximation. The applicability of the
concept of local equilibrium should diminish, however, as a system is driven far from
equilibrium. Nevertheless, questions of "how large drive produces a far-enough
state" and "couldn't one try to find a new equilibrium state near-by" are regularly
asked and have legitimacy. So I will try to illuminate the problems on the example
of the fluctuation-dissipation theorem much discussed nowadays due to attempts
of associating effective temperatures with the various
stages of relaxation in glasses \cite{Tinglass} or with steady states in
granular materials \cite{Tingranu}.

Let us consider a simple system of Ising variables ${\sigma}$ with Hamiltonian
${\cal H}_0$ and in equilibrium at temperature $\beta=1/(k_BT)$. Assuming
that there is an external field $H$ coupling linearly to the macroscopic
magnetization $M=\sum_i\sigma_i$, one can write the equilibrium distribution
function as
\begin{equation}
P_{eq}( \sigma )=Z^{-1} e^{-\beta {\cal H}_0( \sigma )+\beta HM( \sigma )} \,
\label{eqdist}
\end{equation}
The average value of the magnetization is given by
\begin{equation}
\langle M\rangle =Z^{-1} \sum_\sigma M(\sigma )
           e^{-\beta {\cal H}_0( \sigma )+\beta HM( \sigma )} \, .
\label{eqmag}
\end{equation}
and the static limit of the fluctuation-dissipation theorem is obtained as
\begin{equation}
\chi_{_M}=\left .\frac{\partial\langle M\rangle}{\partial H}\right |_{H\to 0} =\beta
            \langle M^2\rangle\,
\label{eqsusc}
\end{equation}
where we assumed the system to be in the high-temperature phase
($\langle M\rangle =0$).
Note the simplicity and the accompanying generality of this derivation.
It uses only the fact that the
external field is linearly coupled to the quantity ($M$) we are considering.

The fluctuation-dissipation theorem is used in many ways.
It helps simplify field-theoretic studies of fluctuations through
diagrammatic expansions and it also gives a powerful
checking procedure in both experiments and Monte Carlo simulations.
Note that eq.(\ref{eqsusc}) can also be used to define
the temperature of the system through
$\beta=\chi_{_M}/ \langle M^2\rangle$,
and the temperature defined in this way would be the same when
using different "$M$"-s and conjugate fields "$H$".

It is clear that a fluctuation-dissipation theorem generalized to
NESS would be extremely useful. Let us now try to imagine how a similar
relationship may arise when we drive the above system away from equilibrium
(e.g. by attaching two heat baths of different temperatures). If the system
relaxes to a NESS then there will be steady-state distribution function, ($P$)
but the effective Hamiltonian ($\ln P$) will contain all the interactions allowed
by the symmetries of the system. Thus, assuming that the effective Hamiltonian
can be expanded in $H$, one finds in the $H\to 0$ limit
\begin{equation}
P_{ne}( \sigma )\sim  e^{-a {\cal H}_1( \sigma )
                          +b H\left[M( \sigma) + S_3(\sigma )+ ...\right]} \,
\label{NESSdist}
\end{equation}
where $S_3$ is a notation for sums over all three-spin clusters with different
couplings for different types of spatial arrangements of the three spins.
Furthermore, $a$, $b$ and all other newly generated couplings depend on the
original couplings in ${\cal H}_0$ and on the temperatures of the heat baths.

Now a derivation of the fluctuation-dissipation theorem similar to the
equilibrium case yields a more complicated equation
\begin{equation}
\chi_{_M}=\left .\frac{\partial\langle M\rangle}{\partial H}\right |_{H\to 0} =
            a\left [ \langle M^2\rangle +\langle MS_3\rangle + ... \right ]\, .
\label{noneqsusc}
\end{equation}
There are two ways a simple form for the fluctuation-dissipation
theorem may emerge from eq.(\ref{noneqsusc}). One is that a nonlinear field
\begin{equation}
Q=M+S_3+...
\label{nonlinfield}
\end{equation}
that is conjugate to $H$ can be introduced (and effectively worked with).
Then one obtains
\begin{equation}
\chi_{_Q}=a\langle Q^2\rangle
\label{nonlinsusc}
\end{equation}
and thus $a$ becomes the nonequilibrium $\beta$.

The other possibility is that a mean-field type decoupling
scheme works well and then
\begin{equation}
\langle MS_3\rangle =\langle M^2\rangle f(C_2)
\label{decoup}
\end{equation}
where $f(C_2)$ is a functional of the
two-point correlations (and similar expressions are obtained for
$\langle MS_{2n+1}\rangle$).
Then equation (\ref{noneqsusc}) becomes
\begin{equation}
\chi_{_M}=a {\cal F}(C_2)\langle M^2\rangle \, .
\label{correlsusc}
\end{equation}
If the theory provides ${\cal F}(C_2)$ then
the effective temperature can again be read of from the above generalized
fluctuation-dissipation relationship.

There are problems with both lines of reasoning. Apart from the practical
difficulties of nonlinear fields and the validity of mean-field
type approaches, the main conceptual difficulty is the fact that changing from
the magnetization to other fields (e.g. energy) the nonequilibrium version
of the fluctuation-dissipation theorem leads to different values for the
same "temperature" \cite{Sollich}. There are cases where the above schemes
generalized to time-dependent processes work both at the theoretical
\cite{{Tinglass},{JLBarrat},{Liu}} and the experimental levels \cite{effTexp}
but there are clear examples when the concept of effective temperature
does not apply \cite{Sollich}. Thus the meaning and use of nonequilibrium
temperature has not been clarified enough to make a verdict on it.

\section{Phase transitions far from equilibrium}
\label{ptffe}
As mentioned in the Introduction, nonequilibrium phase changes constitute
a large part of interesting natural phenomena and
they are studied without worries about wider contexts.
From a general perspective, on the other hand, the investigations of
nonequilibrium phase transitions \cite{{SchmZia},{Marrobook}} can be viewed
as an attempt to understand the robust features of NESS. This view is based on
the expectation that the universality displayed in critical phase transitions
carries over to criticality in NESS as well. If this is true then
studies of the similarities to and differences from equilibrium will lead to
a better understanding of the
role and general consequences of the dynamics generating NESS.

In the following subsections, we shall construct, describe, and discuss models
which display nonequilibrium phase transitions. Apart from getting
familiar with a few interesting phenomena, the main general conclusion
of these discussions should be
that dynamical anisotropies often yield dipole-like effective
interactions \cite{{Schm},{BZ},{uwerev}}
and, furthermore, competing non-local dynamics (anomalous diffusion) generates
long-range, power-law effective interactions \cite{DRT}. Along the way, we shall
also understand that the detailed-balance violating aspects of
local relaxational dynamics do not affect the universality class of
the nonequilibrium phase transitions \cite{{GrinHe},{uwerev}}.

\subsection{Differences from equilibrium - constructing models with NESS}
\label{modelconst}

The violation of detailed balance has the consequence that
not only the interactions determine the properties of the NESS but the dynamics
also plays an important role. In order to understand and characterize the
role of dynamics, a series of simple examples will be discussed in the following
subsections.

First, let us discuss
how to construct a model which yields a NESS in the
long-time limit. A simple way is to attach two heat baths to a system,
each generating a detailed-balance dynamics but at different
temperatures. To see an actual implementation,
let us consider how this is done for the one dimensional
kinetic Ising model. This type of models have been much studied and
a collection of mini-reviews about them can be found in \cite{Privrev}.

The state of the system $\{\s\}\equiv \{\ldots ,\s_i, \s_{i+1}, \ldots\}$
is specified by
stochastic Ising variables $\s_i(t)=\pm 1$ assigned to lattice sites
$i=1,2,\ldots,N$. The interaction is short ranged (nearest neighbor)
$-J\s_i\s_{i+1}$ and periodic boundary conditions
$(\s_{N+1}=\s_{1})$ are usually assumed.
The dynamics of the system is generated by
two heat baths (labeled by $\alpha=1,2$) at temperatures $T_\alpha$, meaning that
the heat baths try to bring the system to equilibrium
at temperature $T_\alpha$ by e.g. spin flips and spin exchanges, respectively.

Let us denote the
rate of the flip of $i$-th spin ($\s_i\to -\s_i$) by
$w_i^{(1)}(\{\s\})$, and let the rate of the exchanges of spins at sites $i$ and $j$
($\s_i\leftrightarrow\s_j$) be $w_{ij}^{(2)}(\{\s\})$.
Then the dynamics is defined by the following
master equation for the probability distribution $P(\{\s\},t)$ :
\begin{eqnarray}
{\partial_t} P(\{\s\},t)
&=\sum_{i}\left[\;
w_i^{(1)}(\{\s\}_i)\; P(\{\s\}_i,t)
-w_i^{(1)}(\{\s\})\; P(\{\s\},t)\; \right]+&
\nonumber \\
&\sum_{ij}\left[\;
w_{ij}^{(2)}(\{\s\}_{ij})\; P(\{\s\}_{ij},t)
-w_{ij}^{(2)}(\{\s\})\; P(\{\s\},t)\; \right]&\label{master}
\end{eqnarray}
where the states $\{\s\}_i$ and $\{\s\}_{ij}$  differ from $\{\s\}$ by the
flip of the $i$-th spin and by the exchange of the $i$-th and $j$-th spins,
respectively.

The assumption that the dynamics is generated by heath baths means that
the rates satisfy detailed balance at the appropriate temperatures:
\begin{equation}
w^{(\alpha)}_{i(j)}(\{\s\})\; P_\alpha^{eq}(\{\s\})=
w_{i(j)}^{(\alpha)}(\{\s\}_{i(j)})\; P_\alpha^{eq}(\{\s\}_{i(j)})
\quad ,
\label{detbal}
\end{equation}
where $P_\alpha^{eq}\sim \exp{[-J/T_\alpha\sum_i\s_i\s_{i+1}]}$ is
the equilibrium distribution of the Ising model
at temperature $T_\alpha$. Eq.(\ref{detbal}) leaves
some freedom in the choice of $w^\alpha$-s,
and one is usually guided by simplicity. The most general spin flip rate
that depends only on neighboring spins has the following form \cite{Glauber}
\begin{equation}
w_i^{(1)}(\s)={1\over 2\tau_1}\left[1-{\gamma\over
2}\s_i\left(\s_{i+1}+ \s_{i-1}\right)\right]\Bigl ( 1+\delta \s_{i+1}
\s_{i-1}\Bigr) \quad .
\label{rate1}
\end{equation}
Without any other heath baths, equations (\ref{master}) and (\ref{rate1})
define the Glauber model \cite{Glauber} which relaxes to the
equilibrium state of the Ising model at temperature $T_1$ defined through
$ \gamma=\tanh (2J/k_BT_1)$. The time-scale for flips is set by $\tau_1$
and $\delta$ is restricted to the interval $-1<\delta<1$.

The competing dynamical process is the generation of
spin exchanges (Kawasaki dynamics \cite{Kawasaki}) by a
second heath bath at a temperature $T_2\not= T_1$. In the
simplest case, the exchanges are between nearest neighbor sites
and the rate of exchange satisfying detailed
balance (\ref{detbal}) is given by
\begin{equation}
w_i^{(2)}(\s)={1\over 2\tau_2}\left[1-{\gamma_2\over
2}\left(\s_{i-1}\s_i+\s_{i+1}\s_{i+2}\right)\right]
\quad .
\label{rate2}
\end{equation}
where $\gamma_2=\tanh (2J/k_BT_2)$ and $\tau_2$ sets the timescale of
the exchanges. It is often
assumed that the exchanges are random ($T_2=\infty$) and thus
$w_i^{(2)}(\s)=1/(2\tau_2)$.

Equations (\ref{master}), (\ref{rate1}) and (\ref{rate2}) define a model which
can be shown to have a NESS and one can start to ask questions about the
phase transitions in this steady state. The generalizations to higher
dimensions, to various combinations competing dynamics (flip - flip,
flip - exchange, exchange - exchange), and other types of dynamical steps
(resulting e.g. from a bulk driving field) should be obvious. Just as it
should be obvious that there are not too many exactly solvable models
in this field and most of the results are coming from simulations
\cite{{SchmZia},{Marrobook}}.

Before turning to results, let us also introduce a Langevin equation description
of competing dynamics. The Langevin approach has been successful in
dynamic critical phenomena \cite{HalpHoh} where the counterparts of the Glauber
and Kawasaki models are called Model A and B, correspondingly. This correspondence
makes the "two-heath-baths" generalization straightforward. The coarse grained
magnetization of the Ising model is replaced by
$n$-component order-parameter field, $S^i({\bf x},t)$, $(i=1,...,n)$
($n=1$ is the Ising model, and $n\to\infty $ is the
spherical limit that allows simple analytic calculations as shown below).
The system evolves under the combined action of local relaxation
(Model A) satisfying detailed balance at temperature $T_1$ and diffusive
dynamics (Model B) at temperature $T_2$ which yields the following
Langevin equation for the Fourier transform $S^i_{\bf q}(t)$
\begin{equation}
\partial_t{S}^i_{\bf q} = -{\cal L}^{(1)}_{\bf q} {S}^i_{\bf q} +
{\eta}^i_1({\bf q},t)
 -{\cal L}^{(2)}_{\bf q}{S}^i_{\bf q} + {\eta}^i_2({\bf q},t) \quad .
\label{genEqmotion}
\end{equation}
Here ${\cal L}^{(\alpha)}_{\bf q} {S}^i_{\bf q}=
\Gamma^{(\alpha)}_{\bf q}\delta F^{(\alpha)}/\delta S_{-\bf q}^i$
with $F^{(\alpha)}$ being the
free energy at temperature $T_\alpha$, and the kinetic coefficient
$\Gamma^{(\alpha)}_{\bf q}$ is enforcing the conservation laws. In particular,
$\Gamma^{(\alpha)}_{\bf q}=\Gamma^{(\alpha)}_0$ in Model A without
conservation laws, and
$\Gamma^{(\alpha)}_{\bf q}=D^{(\alpha)}{\bf q}^2$ for Model B with
diffusive dynamics conserving the total magnetization.
In case $F^{(\alpha)}$ is the Landau-Ginsburg functional, we have
\begin{equation}
{\cal L}_{\bf q}^{(\alpha)} {S}^i_{\bf q} = \Gamma^{(\alpha)}_{\bf q}  \left[
(r_0^\alpha + q^2 ) S^i_{\bf q} +
u \sum_{j=1}^{n}
\int \limits_{\bf q'} \int \limits_{{\bf q}''}
S^j_{{\bf q}'} S^j_{{\bf q}''} S^i_{{\bf q}-{\bf q}'-{\bf q}''} \right]
\label{GLderiv}
\end{equation}
where $r_0^\alpha$ is linear in $T_\alpha$ and $u\sim 1/n$
in the $n\to\infty$ spherical limit. In order to ensure that in case of
a single heat bath, the system relaxes to equilibrium satisfying
detailed balance,
the noise terms in eq.(\ref{genEqmotion} are Gaussian-Markovian random forces
with correlations of the form:
\begin{equation}
\langle {\eta}^i_\alpha ({\bf q},t){\eta}^j_{{\alpha}'}
({{\bf q}'},t')\rangle =
2 \Gamma^{(\alpha)}_{\bf q} \delta_{\alpha {\alpha }'}{\delta}_{ij}
\delta ({\bf q}+{\bf q}') \delta (t-t')  \quad .
\label{Lnoise}
\end{equation}

Eqs. (\ref{genEqmotion}),(\ref{GLderiv}) and (\ref{Lnoise}) define the
model for the particular competing dynamics chosen and we are now ready
to deduce some features of the NESS generated. Of course, just as
in case of kinetic Ising models, the number of possible competing dynamics
is infinite and the question is whether conclusions of some generality
could be reached.
\subsection{Generation of long-range interactions - nonlocal dynamics}
\label{nonlocaldynamics}

The remarkable consequences of competing dynamics can be seen
already on the example of $d=1$ flip-and-exchange model which may
produce ordering even though the interactions are of short range.
Indeed, if $T_1$ temperature spin flips are competing with
$T_2=\infty$ spin exchanges of randomly chosen pairs then
the system orders below a certain $T_{1c}$ \cite{DRT}.

It turns
out that the transition is of mean-field type and this gives a clue
to understanding. Indeed, let us imagine that the rate of spin exchanges
is large compared to the rate of flips. Then the random exchanges
mix the spins in between two flips and the flipping spin sees the "average
spins" in its neighborhood - a condition for mean-field to apply.

The mean field result can also be interpreted as the generation of
infinite-range effective interactions. This interpretation can be put
on more solid base by studying the above model with $T_2=\infty$ spin
exchanges where the probability of exchange at a distance $r$ is
decaying with $r$ as $p(r)\sim 1/r^{d+\sigma}$ (the spins exchanges
are $\sigma$ dimensional Levy flights in dimension $d$). The
system orders again below a $T_{1c}$ and the examination of the critical
exponents reveals \cite{DRT} that the transition is in the universality class
of long-range interactions decaying with $r$ as $J(r)\sim 1/r^{d+\sigma}$.
It is important to note that the above results are nonequilibrium effects
which would disappear if the spin exchanges would also be at $T_1$.

Let us
now see if the same results can be derived from the Langevin equation
approach. The spin flips are translated into the Model A part of the
dynamics while the Levy flights can be represented \cite{Levy-anomdiff}
by anomalous
diffusion with $\Gamma^{(2)}_{\bf q}=D^{(2)}{\bf q}^2$ replaced by
$\Gamma^{(2)}_{\bf q}=D^{(2)}{\bf q}^\sigma$ with $0<\sigma <2$ being
the dimension of the Levy flight. Thus the Langevin equation becomes
\begin{eqnarray}\dot {S}^i_{\bf q} =&- \Gamma_0
( r_0 + q^2 ) S^i_{\bf q}& - {\Gamma}_0 u \sum_{j=1}^{n}
\int \limits_{\bf q'} \int \limits_{{\bf q}''}
S^j_{{\bf q}'}S^j_{{\bf q}''} S^i_{{\bf q}-{\bf q}'-{\bf q}''} +
\eta^i_{\bf q}(t)\hspace{20pt} \nonumber \\
&-Dq^\sigma S^i_{\bf q} + {\bar \eta}^i_{\bf q}(t)&
\, .
\label{Levyeq}
\end{eqnarray}
Note that due to the randomness of the Levy flights ($T_2=\infty$),
the interaction and the nonlinear terms are missing in the
Levy flight part (second line) of the equation. As discussed
in Sec.\ref{modelconst}, the
$\eta$-s are Gaussian-Markoffian random forces with
correlations
$\langle{\eta}^i_{\bf q}(t){\eta}^j_{{\bf q}'}(t')\rangle =
2 {\Gamma}_0 {\delta}_{ij} \delta ({\bf q}+{\bf q}')
\delta (t-t')$ and
$\langle{\bar \eta}^i_{\bf q}(t){\bar \eta}^j_{{\bf q}'}(t')\rangle =
2Dq^\sigma {\delta}_{ij} \delta ({\bf q}+{\bf q}') \delta (t-t')$.

In order to see the generation of long range interaction in the above model,
let us first make an exact calculation \cite{DRT} in the spherical limit
$(n \rightarrow \infty )$ where
fluctuations in $u \sum S^j_{\bf q}(t)S^j_{{\bf q}'}(t)$ may be
neglected and this quantity may be replaced by
\begin{equation}
u \sum_{j=1}^{n} \langle S^j_{\bf q}(t)S^j_{{\bf q}'}(t)\rangle=
unC(q,t)\delta ({\bf q}+{\bf q}') \quad ,
\label{sphericalconstr}
\end{equation}
where the brackets $\langle \rangle$ denote averaging over both the initial
conditions and the noises $\eta$ and $\bar \eta$ (we restrict ourselves to
the study of the high-temperature phase
where the dynamic structure factor
$C(q,t)=\langle S^j_{\bf q}(t)S^j_{-{\bf q}}(t)\rangle $
is independent of $j$).

The decoupling (\ref{sphericalconstr}) leads to a linear equation of motion
and so the self-consistency equation for $C(q,t)$ can be easily derived
\begin{equation}
C(q,t)=2({\Gamma}_0 +Dq^{\sigma} ) \int_{0}^{t} dt'
e^{  - 2\int_{0}^{t'}
\left\{  {\Gamma}_0 [ r_0 + q^2 + unC(q,s)] + Dq^{\sigma}\right\} ds } \, .
\end{equation}
Here, the initial condition $C(q,0)=0$ was used for simplicity. The
$t\to\infty$ limit does not depend on the initial condition
and the equation for the steady state structure factor
$C(q)=C(q,t\to\infty)$ becomes
\begin{equation}
C(q)={ {{\Gamma}_0 + Dq^{\sigma}} \over
         {{\Gamma}_0 (r_0 + q^2 + unS) + Dq^{\sigma}} }
                                               \quad ,
\end{equation}
where $S=\int d {\bf q} C(q)$.

The long-wavelength instabilities are determined by
the $q \rightarrow 0$ form of $C(q)$ which for $0<\sigma <2$
can be written as
$$C(q) \approx (r_0 + \lambda q^{\sigma} + unS)^{-1}
                                               \quad , \eqno(7)$$
with $\lambda =D/ {\Gamma }_0 $. This form coincides with the
long-wavelength limit of the equilibrium structure factor of a
spherical model in which the interactions decay with distance as
$r^{-d- \sigma }$. Consequently, both the self-consistency equation
for $r=r_0 + unS$ and the critical behavior that follows from it
are identical to that of the
equilibrium long-range model. Thus we can conclude that the critical
properties of the NESS are
dominated by an effective long-range potential proportional to
$r^{-d- \sigma }$.

The above conclusion should be valid quite generally
for finite $n$ as well. Looking at eq.(\ref{Levyeq}), one can see that the
correlations in the effective noise $({\eta}_{eff}=\eta + \bar \eta)$
have an amplitude
$2({\Gamma }_0 + Dq^{\sigma })$. One expects that the $Dq^{\sigma }$
term can be neglected in the long-wavelength limit and thus that
the noise $\bar \eta$ in the Levy-flight exchanges can be omitted.
Without $\bar \eta$, however, the system described by equation (2)
satisfies detailed balance and has an effective Hamiltonian which,
apart from the usual short-range interaction pieces,
contains the expected long-range part
$\lambda \int d {\bf q} q^{\sigma} {\sum}_i S^i({\bf q})S^i(-{\bf q})$.

We should note here that $\sigma\to 2$ corresponds to usual diffusion
and that the above arguments changes for $\sigma= 2$  since
no long-range interactions are generated any more, and no change in
critical behavior occurs. This result is another way of saying
that Model A type dynamics is robust against diffusive perturbations
which break detailed balance \cite{GrinHe}. Note also that if both the
competing dynamics are relaxational then, adding up the corresponding
deterministic and noise parts in the Langevin equation (\ref{genEqmotion}),
one can easily deduce that the breaking of the detailed balance
does not change the universality class of the equilibrium
phase transition.

\subsection{Generation of long-range interactions - dynamical anisotropies}
\label{long-range-int}

In order to understand the meaning of dynamical anisotropy, let us
consider the two-temperature, diffusive kinetic Ising model \cite{2tdiff}
on a square lattice. Two heat baths are attached and both of them
generates nearest neighbor spin exchanges.
Exchanges along one of the axis (called  `parallel' direction)
satisfy detailed balance
at temperature $T_{\|}$
while exchanges in the `perpendicular' direction are produced
by a heat bath of temperature $T_{\perp}$.
It is important to note that the interactions $J\s_i\s_j$ are the same along both
axes. It is the dynamics that is anisotropic.

For $T_\perp = T_{\|}=T$, this is the Kawasaki model
\cite{Kawasaki} which relaxes to the equilibrium Ising model at
$T$ and, consequently, it displays a continuous transition.
Since the dynamics conserves the total magnetization,
the ordering for $T< T_c$ appears as a phase separation.

For $T_\perp \ne T_{\|}$, on the other hand,
there is a flow of
energy between the $\|$ and $\perp$ heath baths and the system
relaxes to a NESS. MC simulations
\cite{{2tdiff},{BZ}} show that a critical phase transition is
present for $T_\perp \ne T_{\|}$ as well but
the phase separation is distinct from that occurring in equilibrium.
The interfaces between the domains of up and down spins align with normals
along the directions of lower temperatures.
Thus the symmetries of the ordered states
are different from the symmetry of the
equilibrium order where interfaces with normals
along any of axes coexist (isotropic ordering). As a consequence,
the universality classes of the $\|$ and $\perp$ orderings
are found to be distinct
from the Ising class \cite{{2tdiff},{ShZi92}}.
Renormalization group calculations actually show that the universality class of the
nonequilibrium ordering is that of
a uniaxial ferromagnet with dipolar interactions \cite{Smi}.

A dramatic demonstration of the long-range nature of the interactions
generated by anisotropic dynamics comes from the generalization of the
above model to $n=2$ component spins ($2T-XY$ model). One finds
that the NESS in this system displays an ordering transition \cite{BZ},
a fact that would be in contradiction with the Mermin-Wagner theorem
\cite{Mermin} should the effective interaction be short-ranged.

Let us try now understand the generation of the dipole-like interactions
using the Langevin equation approach and considering the spherical
limit again. The two-temperature, diffusive Ising model
corresponds to competition of two Model B type dynamics along
$\|$ and $\perp$ axes. Thus the equation of motion is
$
\dot {S}^i_{\bf q}(t) = {\cal L}_{{}_\|} (q) {S}^i_{\bf q} +
{\eta}^i_{{{}_\|}}({\bf q},t)
 +{\cal L}_\perp (q){S}^i_{\bf q} + {\eta}^i_\perp ({\bf q},t)
\label{Eqmotion}
$
with the diffusion in the $\alpha ={\|},\perp$ directions described
by the corresponding ${\cal L}_{\alpha}$ terms:
\begin{equation}
{\cal L}_{\alpha} {S}^i_{\bf q} = D_\alpha q_{\alpha}^2  \left[
( r_0^\alpha + q^2 ) S^i_{\bf q} +
u \sum_{j=1}^{n}
\int \limits_{\bf q'} \int \limits_{{\bf q}''}
S^j_{{\bf q}'} S^j_{{\bf q}''} S^i_{{\bf q}-{\bf q}'-{\bf q}''} \right] \, .
\end{equation}
where ${\bf q}=(q_\|,q_\perp)$.
Note the isotropy of the interaction ($q^2$ term) and the different
temperatures ($r_0^\alpha$) for diffusion in different directions.
The noise correlations follow from detailed balance requirements,
$\langle {\eta}^i_\alpha ({\bf q},t){\eta}^j_{{\alpha}'}
({{\bf q}'},t')\rangle \\
=2 D_\alpha q^2_\alpha \delta_{\alpha {\alpha }'}{\delta}_{ij}
\delta ({\bf q}+{\bf q}') \delta (t-t') .
$

Just as in the kinetic Ising model for $r_0^{{}^\|} =r_0^\perp$, we have
Model B with anisotropic diffusion (not a dynamical anisotropy!)
and with an equilibrium steady state, while
a NESS is produced  for $r_0^{{}^\|} \ne r_0^\perp$. The nature
of the phase transitions
in the NESS becomes transparent in the spherical limit
($n\rightarrow \infty$ and $u\sim 1/n$) where the
fluctuations in $u \sum S^j_{\bf q}(t)S^j_{{\bf q}'}(t)$ may be
neglected. This linearizes the equation of motion and allows
to write down a selfconsistency equation for $C({\bf q})=
\langle S^j_{\bf q}(t)S^j_{{\bf q}'}(t)\rangle$. The $t\to\infty$
limit then yields the steady state structure factor in the following form
\cite{BZ}
\begin{equation}
C({\bf q})={
{q^2_{{}_\|} + a q^2_\perp }
\over {q^2_{{}_\|} (r_0^{{}^\|} + q^2 + S) +
          a q^2_\perp (r_0^\perp + q^2 + S) }  }  \quad ,
\label{selfc}
\end{equation}
where $a=D_{\perp}/D_{{}_\|}$ and $S=un\int d {\bf q} C({\bf q})$.

One can see now the origin of dipole-like effective interactions. Because
of the dynamical anisotropy, the
${\bf q}\to 0$ limit of $C({\bf q})$ is different whether first $q_\|\to 0$ and
then $q_\perp\to 0$ or vice versa. This singularity of the long-wavelength
limit translates into such power law correlations in real space
which are characteristic of dipole interactions in equilibrium systems.
Hence the conclusion \cite{{Smi},{BZ}}
that the dynamical anisotropy has generated dipole-like interactions.
Note that this is a nonequilibrium effect. The long-wavelength singularity
disappears as soon as the heat baths have equal temperatures
($r_0^{{}^\|} =r_0^\perp$).

The dynamical anisotropy is a strong effect and its mechanism of
action is rather simple as we have seen above. Accordingly, it
is the most viable candidate to change the universality class of equilibrium
phase transitions by breaking detailed balance \cite{uwerev}.

\subsection{Driven lattice gases, surface growth}

Models of NESS have a long history but the first one that became
the center of attention and was recognized as the "Ising model" of NESS
was the driven lattice gas (see \cite{{SchmZia}} and references therein).
The model can be understood as a kinetic Ising model with the up-spins
being the particles in the lattice gas. Spin exchange
dynamics at temperature $T$ represents the particle diffusion
and an external bulk field $E_x$ drives
the up-spins (particles) along one of the lattice axes $(x)$.
In order to have a NESS with particle current one must also use
periodic boundary conditions in the field direction.

This model displays a critical phase transition in its NESS and
the phase separation that follows at low $T$ is characterized
by strong anisotropy: the interfaces align parallel with $E_x$.
Thus one can see some similarities with the
two-temperature diffusive model of Sec.\ref{long-range-int}.

One can easily recognize that dynamical anisotropy is
at work here. The driving field can be considered as a second heath bath
which generates the essential part of the dynamics along the $x$ direction.
There is a difference,
however, from the two-temperature model of Sec.\ref{long-range-int}
in that the drive now has a directionality (the forward-backward
symmetry of diffusion is broken). As a result the phase transition
is expected to belong to a new (nonequilibrium) universality class
distinct from that of the dipole class. This is indeed what has been obtained
in renormalization group calculations \cite{{SchmZia}}.
Unfortunately, the structure of the long-wavelength singularities
in the two systems are similar and thus there are
difficulties in observing the differences in numerical work.
This problem has generated some debate that is still going on \cite{DDSdebate}.
We believe the debate will not modify the general picture summarized
in \cite{{SchmZia}}, and it will not change
the conclusion about the importance of dynamical anisotropy.

An interesting and important field where even the simplest
systems show "effective" critical behavior
due to the unbounded long-wavelenth fluctuations
is the field of surface growth processes. Most of the roughening transitions
and transitions between various rough phases are genuine nonequilibrium
phase transitions and have been much studied \cite{{Krugreview},{Barabasi}}.
Remarkably, however, these transitions have not provided new insight
into the general features of nonequilibrium criticality, they merely confirmed
that the dynamics and dynamical anisotropy play an important role in
determining the universality classes of growth processes.

One of the unsolved problems that continues to fascinate researchers is the
phase transition in the Kardar-Parisi-Zhang equation
\cite{KPZ}. We shall discuss
the problem of growing surfaces including KPZ equation in connection
with the nonequilibrium distribution functions in Sec.\ref{noneqdist}.

\subsection{Flocking behavior}

Up to this point, we have considered usual physical systems driven
out of equilibrium. Here I would like to give a taste of what awaits one
if the studies are extended to the living realm.

Living creatures can also be viewed as units attached to two heat baths.
One of them is the internal energy source which on a short time-scale
is an infinite bath from which energy can be drawn
at a given rate. The other one is the surroundings to which energy
is lost by dissipation (friction, heat loss, ...)
due to the activity of the unit. This view suggests that
a collection of such self-propelled units will show orderings (nonequilibrium
phase transitions) depending on the interactions between the units, on their
density, on their possible motions and
on the dissipation mechanisms. Indeed, collective behavior
is often observed in flocks of birds, in schools of fish, in swimming cells, etc.
and, as shown below, some of
these phenomena can be described in terms of a surprisingly simple model.

Model \cite{Vicsek-birds}
was introduced to describe the collective motion of self-propelled
particles with birds and bacteria being candidates for these
particles. We shall use the language of "birds" below.

\begin{figure}[hbtp]
	\centering
        \vspace{-0.cm}
	\includegraphics[width=16.5truecm]{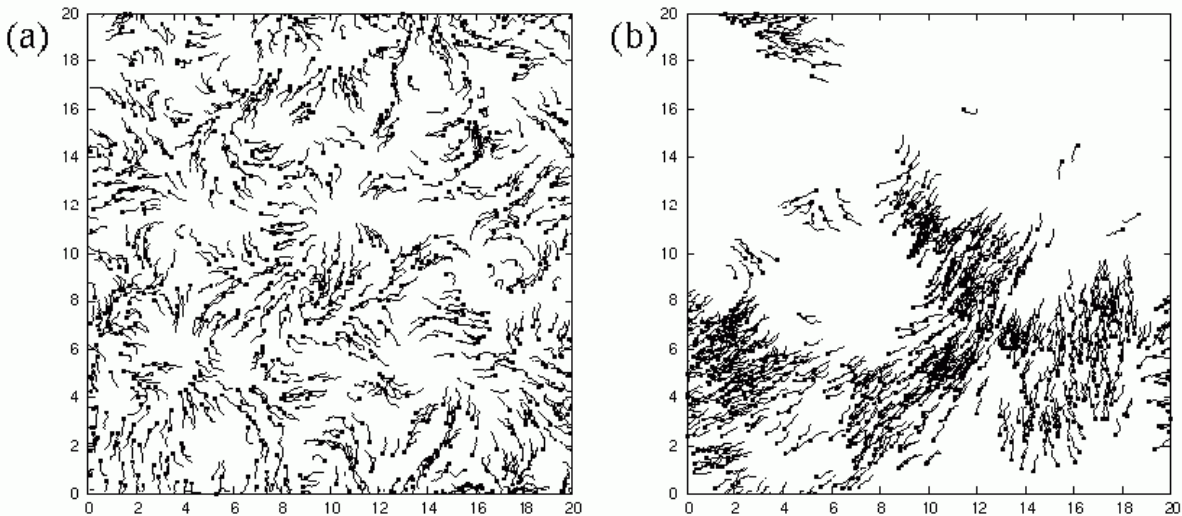}
	\caption[]{Flocking: Trajectories of 10000
        self-propelled particles in the model described in the text.
        The parameters are chosen so that
        the stationary order parameter is $\phi=0.8$. Each particle
is represented by a point marking its current position as well as a
continuous line showing its recent (10 time step long) trajectory. (a) Initial
stage of the relaxation, (b) the stationary regime.
Pictures are curtesy of A. Czir\'ok and T. Vicsek.}
\label{fig:czirok}
\end{figure}
The basic assumptions of the model are that
(i) the birds fly with constant speed $|\vec v_i|=1$ ($i$ is the bird index),
and (ii) the birds adjust their direction $\theta_i$ in time
intervals of $\tau=1$ to the average direction of other birds within a
distance $r$
\begin{equation}
\theta_i(t+\tau)=\langle \theta(t)\rangle_r +\eta_i
\label{reori}
\end{equation}
where $\eta_i$ is random noise with amplitude $\eta$.

Assumption (i) handles the energy in- and outflow by strictly equating them, while
assumption (ii) handles the interactions by seemingly reducing them to interactions
in the space of velocity directions. This is not quite so, however, since the
motion of the birds $\vec x_i(t+\tau)=\vec x_i(t)+\vec v_i(t)\tau$
couples the directional and spatial motions.

The control parameters in the system are $r$, $\eta$ and the density of particles
$\rho$. Keeping $r$ and $\rho$ fixed while varying the "temperature" $\eta$,
one finds that the birds are flying randomly for $\eta>\eta_c(r,\rho)$ while collective
motion develops below $\eta_c$ where the birds tend to move in the same direction.
An order parameter characterizing this spontaneously symmetry breaking
can be chosen e.g. $\phi=|\sum_i^N \vec v_i |/N$. Fig.\ref{fig:czirok}
shows the time evolution deep in the ordered regime ($\phi=0.8$)
starting from a random configuration. One can see that local orientational
order develops in the initial stages of relaxation (the state here shows
resemblance to the states in classical XY ferromagnet)
while the stationary state with almost full orientational order shows
large density fluctuations. The structure and the large density fluctuations
observed in the ordered state
and, furthermore, the measurements of the critical exponents of the transition
\cite{Vicsek-birds} suggest that the ordering in this system is in a
universality so far not encountered.

A remarkable field theory has also been constructed for flocking \cite{Tu}.
It is a generalized Navier-Stokes equation with additional Model A type terms
which drive the velocity to $\langle |\vec v_i|\rangle=1$. This theory
explains the large density fluctuations present in the ordered state.
The investigation of ordering transition is at a higher level of
difficulty, however, and has not been completed yet.

Clearly, much remains to be done before we understand flocking and
before the model can be compared with experiments quantitatively. Nevertheless,
activity is expected in this direction since the model of flocking is
not much more complicated than the more standard NESS models discussed above,
and, at the same time, it has close connection
with experimentally observable,
truly "far-from-equilibrium" phenomena. Hopefully,
by designing and understanding similar models, a kind of "universality map"
of the collective dynamics of self-driven units can be found.

\section{Where do the power-laws come from?}
\label{powlaw}

Systems displaying power law behavior in their
various characteristics (correlation in space
or time, fluctuation power spectra, size-distributions, etc.)
are abundant in nature. The most impressive examples are
found in biology (e.g. the metabolic rate vs. mass relationship
for living creatures displays scaling over 28 decades \cite{West}) but
there are remarkable examples in solid state physics
(power spectra of voltage fluctuations when a current is flowing
through a resistor \cite{Weissman} - 6 decades of scaling),
in geology (the number of earthquakes vs. their magnitude
\cite{earthquakes} - 5 decades) and scaling over 2-3 decades
is seen everywhere [see e.g. the white-dwarf light emission \cite{press},
the flow of sand through an hourglass \cite{shick},
the number of daily trades in the stock market
\cite{mantegna}, water flow fluctuations of rivers \cite{rivers}, the spike trains
of nerve cells \cite{cells}, the traffic flow on a highway \cite{highway},
interface fluctuations \cite{Krugreview},
dissipation in the turbulent systems \cite{turbulence}].

Understanding the (possibly) common origins of scaling in the
above phenomena appears to be a highly nontrivial task.
Power laws, of course, arise naturally in critical phenomena and
we understand them: their origins are in the diverging fluctuations
at the critical point. Thus the first question one may ask is
the following.
\begin{itemize}
\item Can the power laws just be the result of nonequilibrium
      phase transitions and the associated critical behavior?
\end{itemize}
In equilibrium systems, however, one must tune a parameter
to its critical value in order to observe scale-invariant behavior
while nonequilibrium systems appear to be
in scale-invariant states without any tuning. Thus the answer
to the first question appears to be negative.

The wide variety of the phenomena in the above list suggests
that the next question could be as follows.
\begin{itemize}
\item Can the scaling merely be a natural outcome of complex dynamics?
\end{itemize}
After all, we have seen in Sec.\ref{long-range-int} that competing dynamics
may generate long-range (power-law) interactions which may be at the origin of
scaling even away from a critical point. The answer to this question may be
a yes but, unfortunately, many of the problems mentioned are not amenable to
an analysis in terms of simple competing dynamics and then the following question
remains unanswered:
\begin{itemize}
\item What are the ingredients of complex dynamics which determine
the existence and the characteristics (e.g. exponents) of the power laws?
\end{itemize}

There is an attempt to answer all the above question in affirmative
along a logic that begins with
the notion of self-organized criticality (SOC) introduced by Bak, Tang and
Wiesenfeld \cite{BTW}. According to this notion, systems with complex dynamics
tune themselves to a state with a kind of avalanche type dynamics
that is underlying a large number of scale-invariant phenomena.
The notion of SOC has now been
understood in terms of an interplay between local and non-local dynamics
\cite{Grinandothers} which indeed tunes the system \cite{Vespi} to a nonequilibrium
(absorbing-state) critical point. Then the problem of SOC is reduced to
investigating the absorbing state transitions \cite{DickPrivman} and the
characteristics of power laws can be determined by studying the
universality of absorbing state transition. This is an interesting and active field
of research and it is worth understanding the main points.
Accordingly, I will discuss SOC in the
next subsection, and will explain the connection to absorbing state transitions
in the following one.

\subsection{Self-organized criticality (SOC)}

\begin{figure}[hbtp]
	\centering
	\includegraphics[width=10.5truecm]{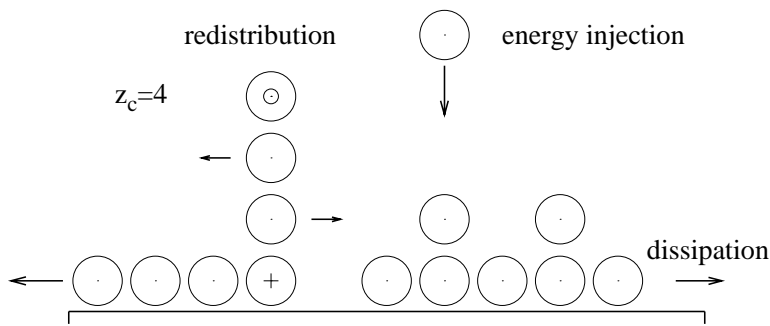}
        \vspace{-0.6cm}
	\caption[]{Sandpile model. Particles (energy packets) are deposited on a
        two-dimensional substrate (one-dimensional section is shown). Injection
        stops when a site becomes active, i.e. it is occupied by more than $z_c$
        particles. Then redistribution
        to neighboring sites take place and particles disappear (dissipation of
        energy) at the edges. The process continues until all active sites are
        eliminated. Then the particles source is switched on again.}
	\label{fig:SOC}
\end{figure}

The first model of SOC was introduced to describe sandpiles.
Later developments, however, made energy packets from
the grains of sand, so the balls in Fig.\ref{fig:SOC}
will be grains at the beginning but will be called energy packets later.
The dynamics defining the model
consists of local and non-local elements. The sites of a (usually) two-dimensional
lattice are occupied by grains and the local aspect of their dynamics is in the
redistribution of the grains. If a site contains more than $z_c$ grains (e.g.
$z_c=4$ on a square lattice) then the site is active and
$z_c$ grains are redistributed to the neighboring
lattice sites. The redistribution of particles (avalanche) continues until
active sites are found.
Clearly, an avalanche stops after a while since the redistribution leads to loss of particles
at the boundaries (or, in terms of the energy model, dissipation occurs at the boundaries).
Once the avalanche stopped, an external supervisor notices it (this is the
nonlocal part of the dynamics) and starts
to add new particles (energy is injected into the system) until a new avalanche starts.

The above dynamics yields a stationary state in the long-time limit, and the steady state
characteristics of avalanches can be measured. Such a characteristics is e.g.
the number of sites $s$ which become
active during the process, and the remarkable feature of this
model is that the distribution of $s$ (and of other quantities such as the spatial size and
the lifetime of the avalanches) is found to display a power law form
\begin{equation}
P(s)\sim s^{-\tau} \, .
\label{aval-dist}
\end{equation}

Thus one discovers that although the model does not contain parameters to tune,
nevertheless it shows critical behavior ($z_c$
can be changed without changing the criticality of the outcome).
This observation generated a large amount of activity and
{\it criticality-without-tuning} was seen in a number of
similar models \cite{SOCrev}. The resulting notion of {\it self-organized criticality}
grew in importance \cite{BakNature} and, accordingly, new effort
was put into understanding how SOC works.

An important feature that was recognized quite early \cite{Grinandothers} is
the existence of a {\it non-local supervisor} who watches the activity of the
avalanches and, upon ceasing of the activity, switches on
the source of particles (or of energy). In principle, non-local dynamics
can generate long-range correlations in both time and space so the
emergence of criticality is not necessary a surprise. Viewing the
problem from another angle,
the non-local dynamics separates the timescales of the avalanches and of the
particle injection. Thus, in practice, there is tuning. Namely, the
system is considered in the limit of particle injection rate going to
zero (actually, the dissipation is also tuned to zero since the particles
disappear only at the boundaries of the system).

The zero injection rate, however, does not have to be a critical point, and
the next important development was \cite{Vespi} the demonstration that it is
indeed a nonequilibrium (absorbing state) critical point.

\subsection{Absorbing state transitions and their connection to SOC}

\begin{figure}[hbtp]
	\centering
        \vspace{0.5cm}
	\includegraphics[width=10.5truecm]{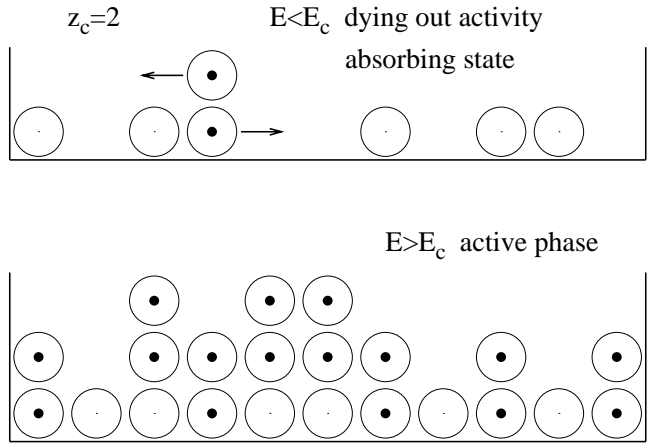}
        \vspace{-0.4cm}
	\caption[]{A one-dimensional fixed energy sandpile model.
        Dynamics is defined by the energy (particles) being
        redistributed if a site contains more than $z_c$ units of
        energy (the units which are redistributed in the next time-step are
        marked by large dots in their center). The total energy of the
        system is conserved since
        the boundaries prevent the loss of energy. }
	\label{fig:fixE}
\end{figure}

Absorbing state transitions appear in many contexts in nonequilibrium
statistical physics \cite{{Marrobook},{Hinrichsen}}, and they are studied
intensively since they are thought to be one of the truly
nonequilibrium phenomena without counterpart in equilibrium systems.
In order to understand the basics of it, let us consider a
fixed energy sandpile model \cite{fixEsand}
shown on Fig.\ref{fig:fixE}.

\begin{figure}[hbtp]
	\centering
        \vspace{-0.1cm}
	\includegraphics[width=10.5truecm]{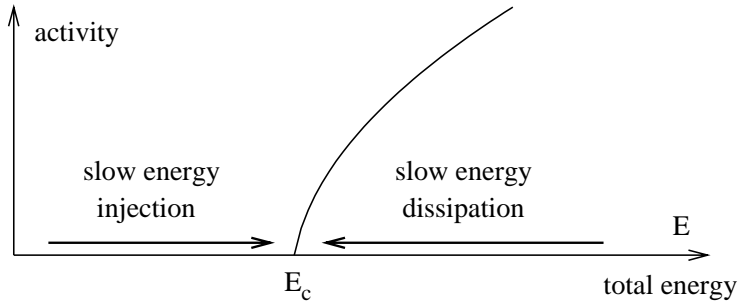}
        \vspace{-0.1cm}
	\caption[]{Activity as a function of the total energy for the
        fixed energy sandpile model described on Fig.\ref{fig:fixE}. The
        evolution of the system as dissipation at the boundaries or
        the energy injection is switched given by the left and right
        arrows, respectively.}
	\label{fig:SOC-abs}
\end{figure}

This model differs from the sandpile model by calling the particles
energy units, and by the absence of both the injection and the dissipation
of energy (particles).
Thus the total energy $E$ is conserved and as one can easily see from
Fig.\ref{fig:fixE} the behavior of the system is essentially different at
small and large values of $E$.
At small $E<E_c$, the activity (redistribution) ceases in the long-time limit
and the system falls into a so called {\it absorbing state}. For large $E>E_c$,
on the other hand, there are always active sites and the redistribution
continues forever. For $t\to\infty$,
the system settles into a steady state which is called
{\it active state} and the {\it activity} can be quantified by measuring e.g. the
number of active sites. One finds then that
the absorbing state transition (i.e. the absorbing-active state transition)
is a critical phase transition with the
{\it activity} changing continously through the transition point $E_c$
(see Fig.\ref{fig:SOC-abs}).

Once the absorbing state transition is understood, it is easy to make
the connection to SOC. Indeed, let us assume that we have the fixed energy
sandpile in the active state ($E>E_c$) and let us switch on the dissipation at the
boundaries. Then the energy decreases slowly (note that the
dissipation is proportional to the surface of the sample while $E$ is
proportional to the volume). This lowering of energy will continue until
$E$ reaches just below $E_c$ when the system falls into the absorbing state
and thus the dissipation stops.

Let us now return to the fixed energy sandpile but this time let us
start from an absorbing state ($E<E_c$) and switch on the "external supervisor"
who is injecting energy into the system.
The supervisor is required to stop the injection if
adding of the last energy packet started activity in the system.
This process increases the energy $E$ infinitesimally
slowly and brings the system near
and perhaps slightly past the threshold of activity $E=E_c$.

Now, if both the dissipation at the boundaries and the "external
supervisor" are present then the fixed energy sandpile model is
nothing else but the
sandpile model generating SOC. And we see that SOC emerges because
the combined action of the dissipation and the "supervisor" brings the
system to the critical point of the absorbing state transition of the
fixed energy sandpile model.

The mechanism unmasked above is rather general and present in many
models of self organized criticality \cite{{Vespi},{Rosso}}. The value of recognizing
this mechanism lies in making it possible to describe and calculate
scaling properties of SOC by studying "usual" nonequilibrium phase transitions.
In particular, one may hope that field-theoretic description of SOC may be
obtained through studies of the appropriate absorbing state transitions.

Of course,
absorbing state transitions are numerous and it is not obvious which one
is in the same universality class as a given system displaying SOC.
In general, continuous phase transitions to an absorbing state are in the
universality class of {\it directed percolation} \cite{{Janssen},{Cardy},{GrassbDP}}
that can be described by the following reaction diffusion process
\begin{equation}
A\to A+A  \quad , \quad A\to 0 \quad , \quad A0 \leftrightarrow 0A \, .
\label{DP}
\end{equation}
Directed percolation is rather robust to various changes in its rules but
the presence of extra symmetries (conservation laws) may change
the universality class of an absorbing state transition. A well
known example is the {\it parity conserving process}
\cite{{GrassbCP},{Menyhard},{CardyCP}} which has the following
reaction-diffusion representation
\begin{equation}
A\to A+A+A  \quad , \quad A+A\to 0 \quad , \quad A0 \leftrightarrow 0A \, .
\label{PC}
\end{equation}
Both of the above processes have been much investigated and the
scaling properties have been accurately determined. Furthermore,
understanding (if not complete solution) has emerged even on
field theoretic level \cite{{Janssen},{Cardy},{CardyCP}}.
Unfortunately, neither of the above processes have been
directly related to models of SOC. Accordingly, the present day research is
concentrated on absorbing state transitions which have more contact
with SOC. An example is the critical point observed in the so called
{\it pair contact process} \cite{PCP} where particles diffuse
only through the birth-death processes given by the reaction scheme
\begin{equation}
A+A\to A+A+A  \quad , \quad A+A\to 0 \, ,
\label{PCP1}
\end{equation}
where the first and the second reactions take place with probabilities
$p$ and $1-p$, respectively. A related problem is the epidemic model
\cite{Pastor} where the reaction scheme
\begin{equation}
A+B\to B+B  \quad , \quad B\to A \quad , \quad B0 \leftrightarrow 0B\, ,
\label{epidemic1}
\end{equation}
describes static healthy subjects ($A$) getting infected by diffusing
infectious agents ($B$) who, in turn, recover with time.

The last two models are close to the fixed energy sandpile models
(and thus to SOC) in that
both of them have an infinite number of absorbing states
and their coarse-grained description involves an order parameter
(the active particles) coupled to a static field (the temporarily
immobile particles). It has recently been suggested
that the similarity may go deeper, i.e. they all  belong to the
same universality class \cite{FvW}. This conclusion is based on
a field-theoretic calculation near dimension $d=6$ \cite{FvW}
using Langevin equations which were suggested on
phenomenological grounds for the processes (\ref{PCP1}) and (\ref{epidemic1})
\cite{{Munoz},{fixEsand}}. At this point there is still a debate about both the
applicability of the Langevin equations and the validity
of the results in lower dimensions. Nevertheless, it appears that
the approach of SOC through absorbing state transitions
may be coming to an interesting and satisfactory conclusion.

Of course, one should not forget that apart from the connection to SOC,
absorbing state transitions
in general constitute an important problem in the theory of NESS.
The field is developing fast and there are many interesting details
scattered across the papers. A guide to the models and to the
extensive literature about them can be found in recent reviews
\cite{{Hinrichsen},{Odor}}.

\section{Distribution functions in nonequilibrium steady states}
\label{noneqdist}

The simplicity of the description of equilibrium system lies in the existence of
the Gibbs distribution i.e. in the elimination of the dynamics from the
calculation of averages. Although dynamics is clearly important in
nonequilibrium steady states, it is not inconceivable that a prescription
exist for a nonequilibrium equivalent of the Gibbs distribution which
would include the essential features of the dynamics. Such a distribution function
may have singularities as shown in simple examples \cite{{TT},{DerrPDF}}
or it may have problems
with the additivity of the associated entropies (which is not unexpected in
systems with long-range correlations) \cite{Tsallis, Superstat}. Nevertheless,
a prescription with well defined restrictions on its applicability
would be valuable and the search for nonequilibrium
distribution function(s) has been going on for some time.

A phenomenological approach to the above problem is the non-extensive statistical
mechanics \cite{Tsallis}, an approach that takes its name from the
nonextensive character of the postulated entropy.
This approach has been much developed during the last decade, and
not surprisingly, is has its success in connection with systems which
have long-range interactions or display (multi)fractal behavior \cite{Tsallis}.

Below we shall present a alternative approach that is somewhat less general
but it is based on the extention of our knowledge of universality
of distribution functions in strongly fluctuating systems.

\subsection{Power laws and universality of nonequilibrium distributions}
\label{dfunc}

Distribution functions of additive quantities such as e.g. the total magnetization
in the Ising model are Gaussian in usual equilibrium systems. This Gaussianity
follows from the central limit theorem that is applicable due to the correlations
being short ranged away from critical points. At critical points, however, the
power law correlations result in non-gaussian distributions.
The emerging distributions are quite restricted in their
possible shapes, however, the reason being that the distribution functions at
critical points are scaling functions and their shape
is determined by the universality class associated with the given
critical point.

The above observation can be used to develop a classification
of nonequilibrium distribution functions. Namely, one knows that "effective"
criticality (i.e. strong fluctuations and power-law correlations) is the norm for
nonequilibrium steady states. Of course, the "effective" critical behavior is
determined not only by the interactions but by the dynamics as well. Accordingly, one
may expect that the scaling functions (and thus the distribution of macroscopic
quantities) are determined by the nonequilibrium
universality classes. Once we build a gallery of such scaling
functions, we can use them in the same way as in the equilibrium
case: We can identify symmetries and underlying mechanisms in experimental systems;
we may find seemingly different systems belonging to the
same universality class, and thus we can discover common underlying
processes present in those systems. We can also
use these distribution functions to find the critical dimension of a model and
the applications are restricted only by imagination. Below we shall show
how to calculate these distribution functions in simple systems and present a
few applications.

\subsection{Picture gallery of scaling functions}

The simplest nonequilibrium systems displaying "effective" criticality
are the growing surfaces \cite{{Krugreview},{Barabasi}}. They are
rough quite generally which means that
the mean-square fluctuations of the surface diverge with system size.
The roughness is defined by
\begin{equation}
w_2=\frac{1}{A_L}\sum_{\vec r} \left[\,h({\vec r},t)-\overline{h}
\,\right]^2 \sim L^\chi \quad ,
\label{widthsq}
\end{equation}
where $A_L$ is the area of the
substrate of characteristic linear dimension $L$,
${\overline h}=\sum_{\vec r}h({\vec r},t)/A_L$
is the average height of the surface, and $\chi$ is a
critical exponent characterizing the given universality class.
We shall be interested in the
steady-state distribution of $P(w_2)dw_2$ and expect
that due to criticality, the diverging scale $\langle w_2\rangle$ will be the
only relevant scale in $P(w_2)$ and, consequently, it can be written in
a scaling from
\begin{equation}
 P_L(w_2)\approx \frac{1}{\langle w_2\rangle_{{}_L}} \Phi \left(
 \frac{w_2}{\langle w_2\rangle_{{}_L}} \right) \quad ,
 \label{Phi}
\end{equation}
where $\Phi(x)$ is a scaling function characteristic of the universality
class the growth process belong to. Below, we show
how to calculate $\Phi$ for a simple growth process (Edwards-Wilkinson
equation \cite{Krugreview}) and will demonstrate that the $\Phi$-s are different for
growth processes distinct in the sense of distinct universality classes.
\begin{figure}[hbtp]
	\centering
        \vspace{0.3truecm}
	\includegraphics[width=9.5truecm]{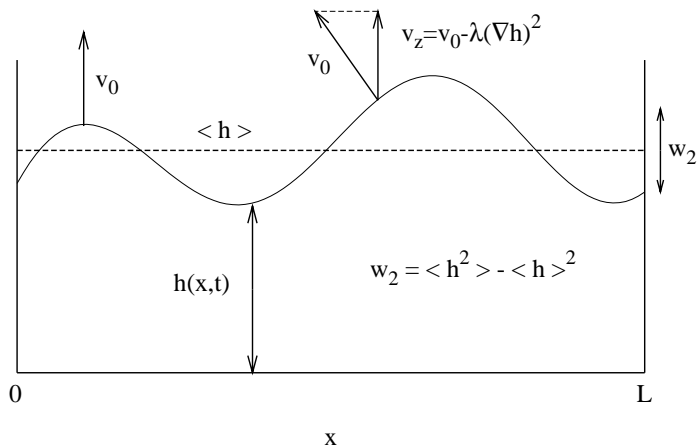}
        \vspace{-0.3cm}
	\caption[]{Surface growth. The height of the surface
        above the substrate is given by $h(x,t)$ and the width of the
        surface $w_2$ is characterized by the mean-square fluctuation.
        The vertical velocity of the surface in general a function
        of the local properties of the surface,
        $v(\partial_xh,\partial_{x}^2h,...)$.}
	\label{fig:surface}
\end{figure}

Let us begin by discussing the equations for
growing surfaces. In general, deposition
of particles on a substrate, under the assumption that the surface
formed is a single valued function
$h(x,t)$,  can be described by the equation
$\partial_t h=v(h)$ where $v(h)$ gives the velocity of advance of the interface
perpendicular to the substrate (Fig.\ref{fig:surface}).
The velocity is usually written as $v(h)=v_0+F(h)+\eta$ where
$v_0$ is the average velocity
due to the average rate of deposition, $F(h)$ is related both to the
motion of the particles on the surface and to the dependence of the
growth on the inclination of the surface. Finally  the fluctuations
in the above processes are collected in $\eta$ which is assumed to be
a Gaussian white noise in both space and time.

A simple form for $F(h)$ follows from the assumption that particles
like to stick at points with large number of neighbors i.e.
at large $\partial_x^2$. Then, $F(h)$ is approximated
as $F(h)=\nu \Delta_x^2$ and, in the frame moving with $v_0$, one has
the Edwards-Wilkinson (EW) model \cite{Krugreview} of surface growth
\begin{equation}
\partial_th(x,t)=\nu \Delta h(x,t) +\eta(x,t) \, .
\label{EWeq}
\end{equation}
This equation can be solved and one finds that the steady-state
probability distribution is given by
\begin{equation}
{\cal P}[h(x)]=Ae^{-\frac{\sigma}{2}\int_0^L(\nabla h)^2dx}
\label{EWhdist}
\end{equation}
where $\sigma$ is related to $\nu$ and to the amplitude of the white noise.

\begin{figure}[hbtp]
	\centering
        \vspace{0.2cm}
	\includegraphics[width=9.5truecm]{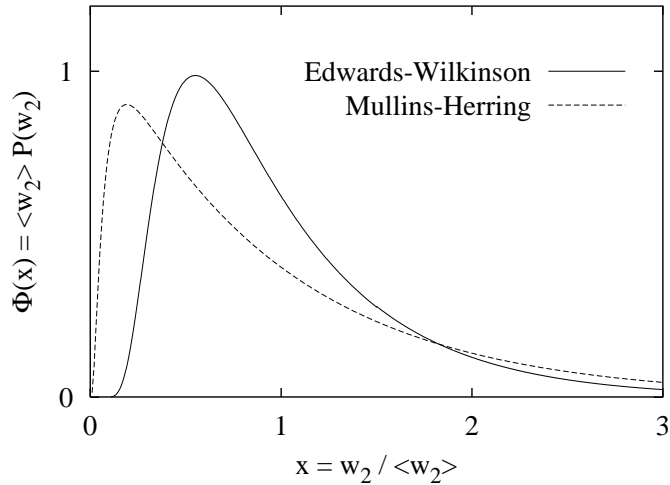}
        \vspace{-0.2cm}
	\caption[]{Comparison of the scaling functions for the EW and MH
        models. }
	\label{fig:EW-MH}
\end{figure}

Once ${\cal P}[h(x)]$ is known, $P(w_2)$ is formally obtained from
\begin{equation}
P(w_2)=\langle \delta(w_2-[\overline{h^2}-\overline{h}^2])\rangle
\label{Pw2definition}
\end{equation}
where the average $\langle \rangle$ is over all $h(x)$ with
the distribution function ${\cal P}[h(x)]$ (note that $\overline {h^n}$ is a
spatial average and it is still a fluctuating quantity).
In practice, it is more convenient to calculate the generating function
\begin{equation}
G(s)=\int_0^\infty e^{-s w_2} P(w_2)dw_2=\langle
e^{-s (\overline{h^2}-\overline{h}^2)}\rangle
\label{Glambda}
\end{equation}
with the above expression demonstrating why $P(w_2)$ can be calculated analytically
in simple models. Namely, if the partition function with $\cal P$ can be found
then the generating function (\ref{Glambda}) is just the partition function
of the model with a quadratic term added and such term usually does not
spoil the solvability of the problem. Indeed, e.g. in case of the d=1 EW model
with periodic boundary conditions,
the problem is reduced to the evaluation of the partition function of a d=1
quantum oscillator thus obtaining \cite{raczdist1}
\begin{equation}
G(s)= \prod_{n=1}^{\infty}
\left(1 + \frac{s L}{\sigma\pi^2 n^2}\right)^{-1} \ .
\label{Glambda2}
\end{equation}
Now one can find the average width diverging
$\langle w_2\rangle =-\partial G(s)/\partial s|_{s=0}=L/(6\sigma)$
in the $L\to\infty$ limit.
Using $\langle w_2\rangle$ to eliminate $L$ from (\ref{Glambda2}), one observes
that $G(s)$ is a function of the product $s\langle w_2\rangle$ only and,
consequently the inverse Laplace transform yields $P(w_2)$
the scaling form (\ref{Phi}). The calculation of the scaling function
$\Phi(x)$ consist of collecting contributions from the poles in $G(s)$
and one obtains
\begin{equation}
\Phi(x) =  \frac{\pi^2}{3}
\sum_{n=1}^{\infty} (-1)^{n-1} \,n^2\,\exp\left(-\frac{\pi^2}{6}n^2x\right) \ .
\label{PhiEW}
\end{equation}
Fig.\ref{fig:EW-MH} shows the above function displaying a characteristic shape
of exponential decay $\Phi(x)\sim e^{-\pi^2x/6}$ at large $x$
and essential singularity $\Phi(x)\sim x^{-5/2}e^{-3/(2x)}$ for $x\to 0$.

On Fig.\ref{fig:EW-MH} we have also included the results for the so called
curvature driven growth process which is also called
the Mullins-Herring model of surface growth \cite{Krugreview}. This is a model
where the rearrangement of deposited particles goes on by surface diffusion
and the particle current $j_h$ is towards places where there are many
neighboring particles i.e. $\Delta h$ is large. This means
that $j_h\sim \nabla \Delta h$ and $F(h)=-\zeta\Delta^2 h$. The resulting equation
is called the Mullins-Herring (MH) equation
\begin{equation}
\partial_th(x,t)=-\zeta \Delta^2 h(x,t) +\eta(x,t) \, .
\label{MHeq}
\end{equation}
The surfaces described by the MH equation belong to a
universality class distinct from that of the EW growth.
Indeed, the MH equation can be solved easily and one finds that
$\langle w_2\rangle_{MH}\sim L^2$
in contrast to the EW result $ \langle w_2\rangle_{EW}\sim L$. Accordingly,
the scaling function should also be different. A calculation similar
to that described above for the EW case verifies this expectation
\cite{PRZ1994} as can be observed on Fig.\ref{fig:EW-MH}.

An important
point to remark about the comparisons of the EW and MH curves is that
they are well distinguishable. Their maximum, their small $x$ cutoff, and
their decay at large $x$ are all sufficiently different so that no ambiguity
would arise when analyzing experimental data. Indeed, the $d=2$
versions of the above $\Phi$-s as well as a number of others characterizing
various growth processes have been obtained in \cite{RP1994} and it did
not appear to be difficult to pick the scaling function which
was corresponding to a given set of experiments \cite{RP1994}.

Finding out the universality class of a growth process is one possible
application if one has a sufficiently developed gallery of scaling functions.
Below we discuss other possibilities for application.

\subsection{Upper critical dimension of the KPZ equation}
The KPZ equation \cite{KPZ} is the simplest nonlinear
model describing growth in terms of a moving interface.
It differs from the EW model by taking into account that
the surface grows in the direction of its normal provided
the incoming particles have no anisotropy in their arrival
direction. Then the $z$ component of the velocity of the surface
has a correction term proportional to $(\nabla h)^2$
as shown on Fig.\ref{fig:surface} and the equation in lowest order
in the nonlinearities becomes the so called KPZ equation
\begin{equation}
\partial_t h = \nu {\vec \nabla}^2 h +
\lambda ({\vec \nabla}h)^2 +\eta \, .
\label{KPZeq}
\end{equation}
Here $\nu$ and $\lambda$ are
parameters, while $\eta({\vec r},t)$ is again a Gaussian white noise.
The steady state surfaces generated by (\ref{KPZeq}) appear
to be rough (critical) in any dimension if the nonequilibrium drive
($\lambda$) is large enough. Since eq.(\ref{KPZeq}) gives account of a number of
interesting phenomena (Burgers turbulence, directed
polymers in random media, etc.)
lots of efforts have been spent on finding and understanding
the scaling properties of its solutions \cite{{Krugreview},{Barabasi}}.
Nevertheless, a number of unsolved issues remain, the question
of upper critical dimension ($d_u$) being
the most controversial one. On one hand mode-coupling and other
phenomenological theories
suggest that $d_u=4$ \cite{modecoup} while all the numerical work
fail to find a finite $d_u$ and the indication is that $d_u=\infty$ \cite{KPZnum}.
Below I would like to show how the scaling
functions of the roughness can shed some light on this controversy \cite{KPZdistr}.

\begin{figure}[hbtp]
	\centering
	\includegraphics[width=10.5truecm]{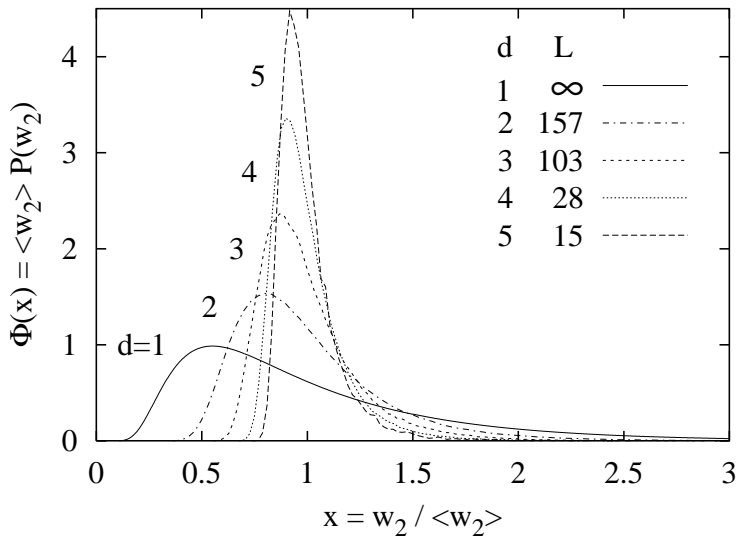}
	\caption[]{Roughness distribution for KPZ steady-state
        surfaces in dimensions $d=1-5$. }
	\label{fig:KPZdistr}
\end{figure}

Let us begin with the observation that scaling functions
do not change above $d_u$. Thus if we build $\Phi(x)$
in dimensions $d=1-5$ and observe that they differ significantly in $d=4$ and $5$
then we can conclude that $d_u> 4$.
Since $\Phi(x)$ cannot be exactly calculated for $d\ge 2$ we must
evaluate it through simulations with the results displayed on
Fig.\ref{fig:KPZdistr}.

As one can see on Fig.\ref{fig:KPZdistr}, the scaling functions
change smoothly with $d$. The $\Phi(x)$-s get narrower and
more centered on $x=1$ with increasing $d$, and there is no
break in this behavior at $d=4$. The
equality of the d=4 and 5 scaling functions appears to be excluded.

Of course, our result is coming from numerical work with the same
general conclusion as in previous studies. So, why
is it more believable? Because one of the main criticism of numerical
studies does not apply to it. Namely, no
fitting parameters and fitting procedures are used in contrast to
the usual determination of
critical exponents. One just builds the histograms, calculates the averages
to determine the scaling variable and plots the scaled histogram. This is
clear and well understood but there is another remarkable
feature in these
scaling functions the origin of which is less obvious. Namely, the scaling
functions in principle should depend on the size of the system,
\begin{equation}
\Phi(x)=\Phi_L(w_2/\langle w_2\rangle_{_L}) \, .
\label{finsizePhi}
\end{equation}
What is observed, however, is that the $L$ dependence is practically
all in $\langle w_2\rangle_{_L}$ and $L$ dependence of the shape of $\Phi$
(explicit dependence on $L$) disappears already at small $L$.
These are important points and the KPZ application of the scaling functions
was mainly chosen to emphasize them.

It seems that these scaling functions are versatile tools which can be used
in computer science \cite{Korniss} as well as in understanding the propagation
of chemical fronts \cite{Tripathy}.
An interesting application was e.g.
the establishment of a connection between the energy
fluctuations in a turbulence experiment
and the interface fluctuations in the
$d=2$ Edwards-Wilkinson model \cite{BHP98}, and
thus prompting a search for an interface interpretation
of the dissipative structure in the turbulent system \cite{Goldenfeld}.
In another case, it helped to make a link between the much studied
$1/f$ noise and the extreme value statistics \cite{1fEVS}. Since the
effective criticality is a real feature of many nonequilibrium system,
we expect that many more use will be found for the scaling functions
discussed in this section.

\section{Quantum phase transitions}
\label{Quantumpt}

Quantum critical points are associated with the change of the symmetry of the
ground state of a quantum system as
the interactions or an external field (control parameter) are varied.
They have been much investigated in recent years with the motivation
coming from solid state physics \cite{Sachdevbook}.
Namely, the strongly-correlated electron systems often
produce power law correlations, and the origin of the observed scale-invariance
is suggested to be
the presence of a quantum critical point at $T=0$ provided the effect of
the quantum phase transition is felt at finite $T$ as well.

From our point of view, the quantum phase transitions are interesting because
they are good candidates for studying the effects of a nonequilibrium drive on
well established
symmetry-breaking transitions. The advantage of these systems is that there is no
arbitrariness in their dynamics (it is given by quantum mechanics),
the one-dimensional
systems are simple with examples of exactly solvable models displaying
genuine critical phase transitions (see e.g. the transverse Ising chains
discussed below) and, furthermore, there is much previous work to build on.

The only problem is how to force a quantum system into a non-equilib\-rium steady state.
An obvious way is to attach two heat baths of different temperatures at the two
ends of a spin chain. Unfortunately, this makes the problem unsolvable (even numerically)
for any reasonable size system \cite{Saito} and thus it is practically impossible e.g.
to draw conclusions about the
long-range correlations generated in the system. Below we show
a way to avoid the problem of heat baths. The idea is that the nonequilibrium
steady states
always carry some flux (of energy, particle, momentum, etc.).
Thus a steady state that is
presumably not very far from the one generated by boundary conditions may be constructed
by constraining the quantum system to have a flux equal to the one generated by
the boundary conditions. For example, in the case
of the transverse Ising chain treated below, we shall constrain the
system to carry an energy
current and will investigate the correlations in this constrained
state.\footnote{It should be noted that this section was not discussed
during the main lectures of the course. It was described only in a seminar for interested students.}

\subsection{Spin chains with fluxes}

As a simple model with critical phase transition, we consider the
$d=1$ Ising model in a transverse field $h$ which has the following Hamiltonian:
 \begin{equation}
 \hat H_I=-\sum_{\ell=1}^N \left(\sigma^x_\ell \sigma^x_{\ell+1}
 + \frac{h}{2} \sigma^z_\ell \right) \, .
 \label{trisinghami}
 \end{equation}
Here the spins $ \sigma^\alpha_\ell$ ($\alpha =x,y,z$)
are represented by $1/2$ times the Pauli matrices
located at the sites
$\ell=1,2,...,N$ of a one-dimensional periodic chain
($\sigma^\alpha_{N+1}= \sigma^\alpha_1$). The
transverse field, $h$, is measured in units of the Ising coupling, $J$, which
is set to $J=1$ in the following.

This model can be solved exactly \cite{{LSM},{Pfeuty}} and it is known
that a second order phase transition
takes place in the system as $h$ is decreased. The
order parameter is the expectation value $\langle \sigma_x\rangle$ i.e.
$\langle \sigma_x\rangle=0$ for $h> 1$, while $\langle \sigma_x\rangle\not=0$ for $h< 1$
and $h_c=1$ is a critical point. The scaling behavior at and near $h_c$ belongs
to the $d=2$ Ising universality class.

In order to constrain the above
system to carry a given energy flux $J_E$ we shall use the Lagrange
multiplier method. Namely, we add a term $\lambda \hat J_E$ to the
Hamiltonian where $\hat J_E$ is the local energy flux operator summed
over all sites, and find that value of $\lambda$ which produces a
ground state with the expectation value $\langle\hat J_E\rangle = J_E$.

The above scheme requires the knowledge of the local energy current, $\hat J_\ell$.
It can be obtained using the quantum
mechanical equation of motion for the energy density $\dot \varepsilon_\ell=
i/\hbar [\hat H_I,\varepsilon_\ell]$, and representing the
result as a divergence of the energy current $\dot \varepsilon_\ell=J_\ell-J_{\ell+1}$.
The calculation yields ($\hbar =1$ is used in the following)
\begin{equation}
\hat J_\ell = \frac{h}{4}\sigma^y_\ell(\sigma^x_{\ell-1}-\sigma^x_{\ell+1})
\label{current}
\end{equation}
and this allows to construct the `macroscopic'
current $\hat J_E=\sum_\ell \hat J_\ell$.  Adding it to $\hat H_I$
with a Lagrange multiplier, $-\lambda$,
\begin{equation}
\hat H =\hat H_I -\lambda \hat J_E \quad .
\label{ham}
\end{equation}
we obtain the Hamiltonian whose ground states with $\langle\hat J_E\rangle = J_E\not=0$
will give us information about the current carrying states of $\hat H_I$.

In order to avoid confusion, we emphasize
that the energy current, $\hat J_E$, is associated with $\hat H_I$
and not with the new Hamiltonian, $\hat H$.
We also note that $\hat H$ is just another equilibrium Hamiltonian,
it differs from $\hat H_I$ by an extra term which breaks the left-right
symmetry of $\hat H_I$.
Finding the ground state of $\hat H$, however, gives us the minimum energy
state of $\hat H_I$
which carries an energy current, $J_E =\langle \hat J_E \rangle$.
Thus the ground-state
properties of $\hat H$ provide us with the properties of the nonequilibrium
steady states of the transverse Ising model.

It turns out that $[\hat H_I,\hat J_E]=0$ and $\hat H$ can be diagonalized by
the same transformations which diagonalize
$\hat H_I$ \cite{ARS97}, and one arrives to a system of free fermions
with a spectrum of
excitation energies given by $\omega_q=\left|\Lambda_q\right|$ where
\begin{equation}
\Lambda_q=\frac{1}{2} \sqrt{1+h^2+2h\cos q} + \frac{\lambda h}{4}\sin q \quad .
\label{excitations}
\end{equation}
with the wave numbers restricted to
$-\pi\le q\le\pi$ in the thermodynamic limit ($N\rightarrow \infty$).
\begin{figure}[hbtp]
	\centering
	\includegraphics[width=9.5truecm]{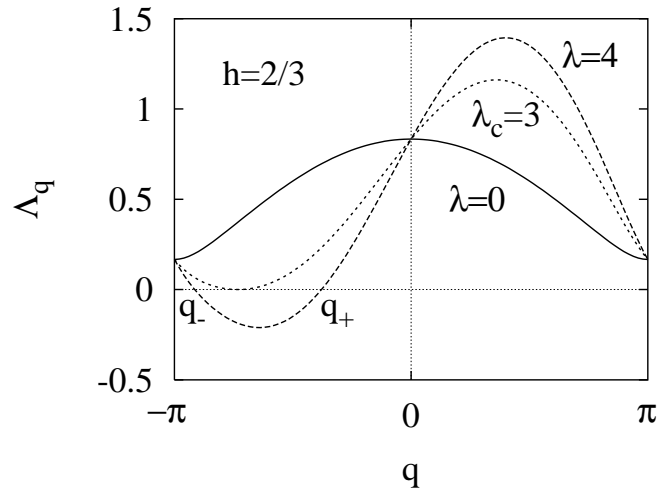}
        \vspace{-0.2cm}
	\caption[]{Spectrum of the transverse Ising model
in the presence of a field ($\lambda$) which drives the current of energy.
The excitation energies are given as $\omega_q=|\Lambda_q|$. Increasing the
drive makes
the ground-state change at a critical $\lambda=\lambda_c$ ($\lambda_c=3$ for
$h=2/3$) when negative energy states start to appear and get occupied.
The qualitative picture is the same at all transverse fields
$h$. }
	\label{spectrum}
\end{figure}

Fig.\ref{spectrum} displays the spectrum
for $h=2/3$ and various $\lambda$ and one can see that
the $q\rightarrow -q$ symmetry of the spectrum is broken for $\lambda \not= 0$.
Nevertheless, for small $\lambda$ the ground-state remains that of the
transverse Ising model ($\lambda=0$) since $\Lambda_q\ge 0$ and
the occupation number representation of the ground state does not change.
Accordingly, no energy current
flows $(J_E=0)$ for $\lambda <\lambda_c$. This rigidity of the ground state
against the symmetry-breaking field which drives the energy current is a
consequence of the facts that the fermionic spectrum of the
transverse Ising model has a gap and
that the operator $\hat J_E$ commutes with $\hat H_I$ (similar rigidity is
observed in the studies of energy flux through transverse $XX$ chain \cite{ARRS1}).
\begin{figure}[hbtp]
	\centering
        \vspace{-0.3cm}
	\includegraphics[width=9.5truecm]{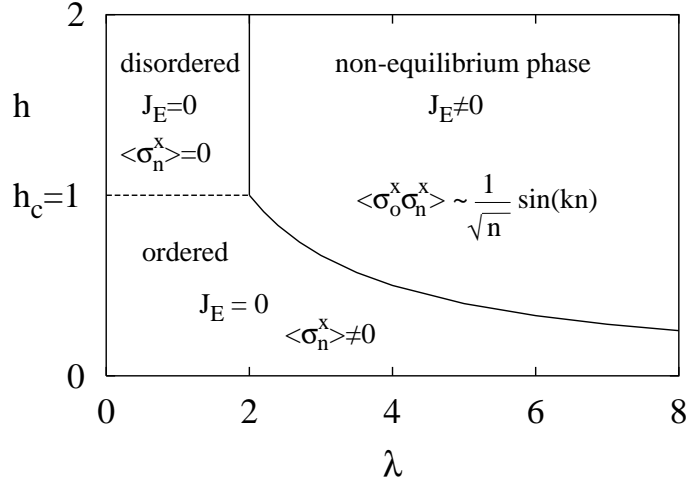}
        \vspace{-0.2cm}
	\caption[]{Phase diagram of the driven transverse Ising model
        in the $h-\lambda$ plane where $h$ is the transverse field while
        $\lambda$ is the effective field which drives the flux of energy.
        Power-law correlations are present in the nonequilibrium phase
        ($J_E\not= 0$) and on the Ising critical line in the
        equilibrium phase ($J_E=0$, dashed line). }
	\label{fig:phasedia}
\end{figure}

The ground-state properties do change when $\Lambda_q<0$ in an interval
$[q_-,q_+]$ and these $q$ states
become occupied. Due to the resulting asymmetry in the occupation of the
$q$ and $-q$ states, the energy current becomes nonzero. The line $\lambda_c(h)$
which separates the region
of unchanged transverse Ising behavior from the $J_E\not=0$ region is
obtained from the conditions $\Lambda_q=0$ and $\partial \Lambda_q/\partial q =0$,
and is displayed on the phase diagram (Fig.\ref{fig:phasedia}) as a solid line.
Another phase boundary on Fig.\ref{fig:phasedia} is shown by dashed line.
It separates the
magnetically ordered ($h<1$, $\lambda<2/h$) and disordered ($h\ge 1$,
$\lambda<2$) transverse Ising regions.
Since the ground state is independent of $\lambda$ for $\lambda<\lambda_c$,
one has the same second order transition across the dashed line
as at $h=1$ and $\lambda=0$
i.e. it belongs to the $d=2$ Ising universality class \cite{Pfeuty}.

Clearly, one can view the region $\lambda<\lambda_c$ as an equilibrium
phase while the $\lambda>\lambda_c$ region as a nonequilibrium one
since there is a nonzero energy flux through the latter. This flux can actually be
calculated easily with the simple result
\begin{equation}
j_E =\langle \hat J_E/N \rangle=
(4\pi)^{-1}\sqrt{\left(1-4/\lambda^2\right)
\left(h^2-4/\lambda^2\right)} \quad .
\label{J}
\end{equation}

Apart from the fact that $J_E\not=0$, the $\lambda>\lambda_c$ region
should also be considered as a distinct phase since the long-range
magnetic order existing for $h<1$ breaks down when $J_E\not=0$ and
the magnetic correlations become oscillatory with
amplitudes decaying as a power of distance. Indeed,
this can be seen by investigating the
$\langle \sigma^x_\ell \sigma^x_{\ell+n}\rangle$
correlations which can be expressed through Pfaffians \cite{Baruch}
and thus making possible numerical calculations for $n\le 100$.
In the presence of long-range order one should have
$\langle \sigma^x_\ell \sigma^x_{\ell+n}\rangle\rightarrow
\langle \sigma^x_\ell \rangle^2 \not=0$ for $n\rightarrow \infty$
while we find that the correlations decay to zero at large distances as
\begin{equation}
\langle \sigma^x_\ell \sigma^x_{\ell+n}\rangle
\sim\frac{Q(h,\zeta)}{\sqrt n}\cos(kn)
\label{xxcorr}
\end{equation}
where the wavenumber, $k=\arccos{(2/\lambda h)}$.
The above result (\ref{xxcorr}) is coming from numerics and it is exact
in the $\lambda\rightarrow\infty$ limit where
the correlations are those of the $d=1$
$XX$ model \cite{Baruch}.

One can observe power-law correlations for $\lambda >\lambda_c$
in other physical
quantities as well. For example, the envelopes of both
$\langle \sigma^z_\ell\sigma^z_{\ell+n}\rangle$ and
$\langle \hat J_\ell \hat J_{\ell+n}\rangle$ correlations
behave  as $n^{-2}$  in the large $n$ limit \cite{ARS97}.
Thus we arrive to the main conclusion of this section, namely
that a simple, exactly soluble quantum system shows power-law correlations
in the current carrying state
in agreement with the notion that power-law correlations are a
ubiquitous feature of nonequilibrium steady states.

Actually, remembering that power-law correlations in quantum models are associated with
a gapless excitation spectrum, we can reformulate the
transverse Ising model result to
see a general connection between the emergence of power-law
correlations and the presence of a current. Indeed, let us
assume that a system with Hamiltonian $\hat H_0$ has a spectrum with a gap
between the ground-state and the lowest excited state. Furthermore, let
$\hat J$ be a `macroscopic' current of a conserved quantity
such that $[\hat H_0,\hat J]=0$. Generally, there is no current in
the ground state and adding $-\lambda\hat J$ to $\hat H_0$
does not change the $\langle \hat J \rangle =0$ result for small $\lambda$.
Current can flow only if some excited states mix with the ground state
and, consequently,
a branch of the excitation spectrum must come down and intersect the
ground-state energy in order to have $\langle \hat J \rangle \not= 0$. Once
this happens, however, the gap disappears and one can expect power-law
correlations in the current-carrying state.
Admittedly, the above argument is not strict and is just a reformulation
(in general terms) of what we learned from the transverse Ising model.
We believe, however, that the above picture is robust and suggestive enough to
try to find other soluble examples displaying the flux $\rightarrow$
power-law-correlations relationship.

\subsection{Quantum effective interactions}

When using the Lagrange multiplier method,
one assumes that the flux generated by boundary conditions can be replaced by
the effective interactions contained in an appropriately chosen global
flux $\hat J$. These interactions are generally short ranged
since the flux is usually a sum of local terms. The short-range
nature of the effective interactions is actually not in contradiction
with the power-law correlation being generated since the Lagrange multiplier
gets tuned in order to achieve a given flux and, furthermore, the tuning is
not quite trivial since $\lambda$ must be increased past a critical value
in order to have e.g. a nonzero flux energy.
The question nevertheless arises whether adding a flux term in $\hat H$ was an
adequate description for the
nonequilibrium steady state which is expected to display power-law correlations.

In order to investigate the above question one would have to solve the
problem with the boundary drive but, as discussed above, an exact solution
does not seem to be feasible. Instead, however, one can prepare initial
conditions which will lead in the long-time limit to a steady flux.
Then the steady state obtained in this natural way can be compared to the one found
by the Lagrange multiplier method. This program has been carried out \cite{ARRS2}
for the $XX$ chain defined by the following Hamiltonian
 \begin{equation}
 \hat H_{XX}=-\sum_{\ell=1}^N \left(\sigma^x_\ell \sigma^x_{\ell+1}+
      \sigma^y_\ell \sigma^y_{\ell+1}  + h \sigma^z_\ell \right) \, .
 \label{trxyhami}
 \end{equation}
In this model, the transverse magnetization $M_z=\sum_i \sigma_i^z$ is
conserved and one can investigate the nonequilibrium states which
carry a given magnetization flux by using the Lagrange multiplier method
 \cite{ARRS1}. At the same time,
the model is simple enough so that one can solve the time dependent
problem where a steady magnetization flux is achieved by
starting with an inhomogeneous initial state that is the ground state at fixed
magnetization but with $m=\langle s^z_n\rangle$ reversed from $m_0$
for $n\le 0$ to $-m_0$ for $n>0$.
\begin{figure}[hbtp]
	\centering
        \vspace{-0.4cm}
	\includegraphics[width=10.5truecm]{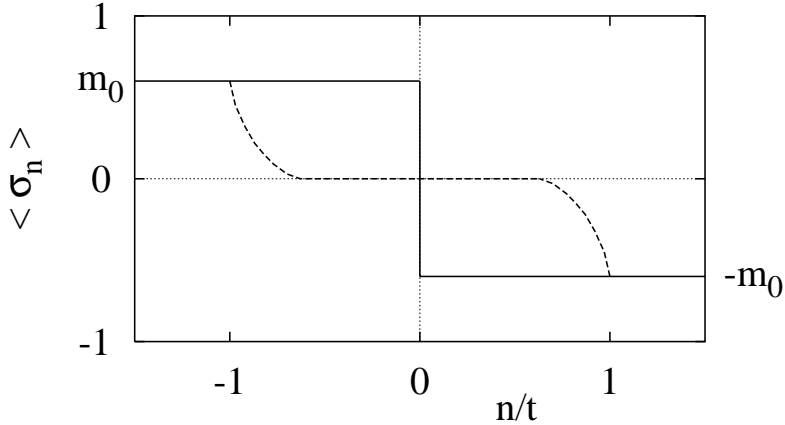}
	\caption[]{Time evolution of the magnetization profile starting
        from a step-like initial conditions shown as solid line. There are
        two fronts going out to $\pm \infty$. They diminish the magnetization
        and leave behind a homogeneous $\langle s^z_n\rangle=0$ state.
        In the scaling limit $t\to \infty$, $n/t\to x$, the
        magnetization $m(n,t)\approx \Phi(n/t)$ is given by
        $\Phi(x)=m_0$ for $-1< x$,
        $\Phi(x)=m_0-\pi^{-1}\arccos(x)$ for $-1<x<-\cos(\pi m_0)$,
        $\Phi(x)=0$ for $-\cos(\pi m_0)< x<0$, and $\Phi(x)=-\Phi(-x)$ for $x>0$
         (dashed line). }
	\label{fig:magprof}
\end{figure}
The time-evolution of this step-like initial state can be
followed exactly and the magnetization profile emerging in the long-time
limit is shown on Fig.\ref{fig:magprof}. The remarkable feature of this
magnetization profile is the middle part which is an $m=0$ homogeneous state
carrying a magnetization flux $j(m_0)$.
Comparing this state to the one generated by adding the flux
term to the Hamiltonian and fixing the Lagrange multiplier to have
the same $j(m_0)$, we find that various expectation values such as
the energy, the occupation number in fermionic representation are all
equal in the two states. Thus the Lagrange multiplier yields a correct
description of the states carrying a magnetization flux.

It should be noted, however, that recent
calculation with an inhomogeneous initial state of different temperatures
($T_1$ for $x<0$ and $T_2$ for $x>0$) has yielded a different result for
the asymptotic state carrying an {\it energy} flux \cite{Ogata01}.
Namely, it was shown that, at least in the neighborhood of $x=0$ and in
the $t\rightarrow \infty$ limit,
the properties of the flux-carrying state can be interpreted in terms of
the ground state of an effective Hamiltonian
\begin{equation}
 \hat H_{\rm eff}=\hat H_{XX}+\sum_{n=1}^N \mu_n \sum_{j=1}^N \hat Q_j^{(n)} \, ,
 \label{jehami}
 \end{equation}
where $Q_j^{(n)}$ is a product of local operators at sites $j$ and $j+n$,
and the interaction is of long-range type since $\mu_n\sim 1/n$ \cite{Ogata01}
(remarkably, the first two operators $\hat Q_j^{(1)}$ and $\hat Q_j^{(2)}$
are those appearing in the Lagrange multiplier treatment of the energy flux in the
$XX$ chain).
Although the homogeneity of the asymptotic state was not shown and thus the comparison
may be questioned, the above result indicates that the Lagrange multiplier
approach may be only a first approximation in describing the flux-carrying states.

In summary, the studies of quantum systems described above strengthen the
view that fluxes
generate long-range correlations. Furthermore, the quantum systems also
give a simple picture of how the emergence of these correlations is
related to the closing a gap in the excitation spectrum above the ground state.

\section{Outlook}
\label{Outlook}

There are topics which are important but were not discussed
in these lectures. To mention a few, there is a large body of work on
one-dimensional systems displaying nonequilibrium phase transitions, on orderings
of granular gases under shear, on pattern formation
with phase transitions described by the complex-coefficient
Landau-Ginzburg equation, and the list could be continued.
My choice of topics mainly reflects my past work and my attempts
to develop simple starting points for making inroads into the
beautiful but rather difficult field of far from equilibrium phenomena.

Finally, I was asked to provide entertainment for readers by
trying to guess the future developments in connection with
nonequilibrium orderings. Well, one of the present problem of the
field is the lack of simple experimental systems which
can be be taken far enough from equilibrium and compared to elementary
models of NESS. Search for such systems will
intensify and I expect that there will be a shift towards biological
problems. There the condition of being far from equilibrium is satisfied
and there may be surprisingly simple phenomena under the guise of
complicated pictures. This line of research may in the future
meet up with game theories generalized to take into account spatial
structures.

Search for better understanding of the emerging effective interactions
will also continue, just as the
sorting out of the absorbing-state transitions (surprising connections
may be still found there, in addition the existing one to SOC).
I also believe that theory of nonequilibrium
distributions will
be much developed, and limiting distributions such as
the ones emerging in extreme statistics will have a much wider use
in physics. These are more or less safe bets. And then there is
the unpredictable part of future.

\acknowledgements
My research partially described in this lecture series has been supported by the Hungarian Academy
of Sciences (Grant No. OTKA T029792).
I thank the organizers for providing ideal surroundings for delivering
these lectures. Thanks are also due to M. Droz and F. van Wijland for helpful
discussions and comments on the manuscript.
\end{document}